# Self-Similarity of ICME Flux Ropes: Observations by Radially Aligned Spacecraft in the Inner Heliosphere

**S. W. Good[1], E. K. J. Kilpua[1], A. T. LaMoury[2], R. J. Forsyth[2], J. P. Eastwood[2], and C. Möstl[3]**

[1]Department of Physics, University of Helsinki, Helsinki, Finland

[2]The Blackett Laboratory, Imperial College London, London, UK

[3]Space Research Institute, Austrian Academy of Sciences, Graz, Austria

Corresponding author: Simon Good (simon.good@helsinki.fi)

**Key Points:**

- Eighteen interplanetary flux ropes observed by radially aligned spacecraft in the inner heliosphere have been examined
- Many of the flux ropes showed significant self-similarities in magnetic field structure at the aligned spacecraft
- Macroscale differences in the magnetic field profiles are consistent with the flux ropes displaying different axis orientations

## Abstract

Interplanetary coronal mass ejections (ICMEs) are a significant feature of the heliospheric environment and the primary cause of adverse space weather at the Earth. ICME propagation, and the evolution of ICME magnetic field structure during propagation, are still not fully understood. We analyze the magnetic field structures of 18 ICME magnetic flux ropes observed by radially aligned spacecraft in the inner heliosphere. Similarity in the underlying flux rope structures is determined through the application of a simple technique that maps the magnetic field profile from one spacecraft to the other. In many cases, the flux ropes show very strong underlying similarities at the different spacecraft. The mapping technique reveals similarities that are not readily apparent in the unmapped data and is a useful tool when determining whether magnetic field time series observed at different spacecraft are associated with the same ICME. Lundquist fitting has been applied to the flux ropes and the rope orientations have been determined; macroscale differences in the profiles at the aligned spacecraft may be ascribed to differences in flux rope orientation. Assuming that the same region of the ICME was observed by the aligned spacecraft in each case, the fitting indicates some weak tendency for the rope axes to reduce in inclination relative to the solar equatorial plane and to align with the solar east-west direction with heliocentric distance.

## Plain Language Summary

Coronal mass ejections (CMEs) are large eruptions of magnetic field and plasma from the Sun. When they arrive at the Earth, these eruptions can cause significant damage to ground and orbital infrastructure; forecasting this 'space weather' impact of CMEs at the Earth remains a difficult task. The impact of individual CMEs is largely dependent on their magnetic field configurations, and an important aspect of space weather forecasting is understanding how CME field configuration changes with distance from the Sun. We have analyzed the signatures of 18 CMEs observed by pairs of lined-up spacecraft and show that their basic magnetic field structures often display significant self-similarities, i.e., they do not often show significant reordering of field features with heliocentric distance. This similarity points to the general usefulness of placing spacecraft between the Sun and Earth to act as early-warning space weather monitors. CME signatures observed at such monitors would likely be similar to the signatures subsequently arriving at the Earth and could be used to produce space weather forecasts with longer lead times.

## 1 Introduction

Coronal mass ejections (CMEs; e.g., Webb and Howard, 2012) are large-scale eruptions of magnetized plasma from the solar atmosphere and one of the most vivid examples of the Sun's dynamism. Their counterparts beyond the corona, interplanetary coronal mass ejections (ICMEs; e.g., Kilpua et al., 2017a), cause significant perturbations within the heliospheric environment (e.g., Möstl et al., 2012), and they are the primary drivers of magnetospheric activity at the Earth (e.g., Kilpua et al., 2017b). How CMEs form, how they are structured, and how they propagate and evolve in interplanetary space are all important considerations for space weather forecasting (Manchester et al., 2017). ICME magnetic field structure is of particular interest due to its central role in solar wind-magnetosphere coupling (Dungey, 1961; Pulkkinen, 2007).

The paradigm of the CME as a twisted flux tube with a relatively well-ordered magnetic field that propagates into the heliosphere as an ICME is well established. These twisted flux tubes, or flux ropes, typically display low variance magnetic fields with a field direction smoothly varying over ~1 day when observed by spacecraft at 1 AU; if such flux ropes also exhibit low plasma $\beta$ and low proton temperatures, they may be described as 'magnetic clouds' (Burlaga et al., 1981). Although many ICMEs are not observed to have clear flux rope signatures (e.g., Cane & Richardson, 2003), those with a rope structure are the usual source of major geomagnetic disturbances (e.g., Wu & Lepping, 2011) since flux ropes can cause sustained periods of high-magnitude, southward magnetic field to be incident at Earth. The survival of CME flux ropes to 1 astronomical unit (AU) and beyond indicates that they hold some degree of structural robustness and stability (Burlaga, 1991; Cargill and Schmidt, 2002).

CME flux ropes change significantly in size, velocity and orientation during propagation, and their global shape can become significantly distorted relative to their initial configuration. The biggest changes in CME velocity, orientation and propagation direction are thought to occur in the corona (e.g. Vourlidas et al., 2011; Isavnin et al., 2014; Kay et al., 2017), while changes in size and shape occur more dramatically in interplanetary space. Changes to CME morphology in the heliosphere are largely driven by interactions with the ambient solar wind. The fall in solar wind pressure with heliocentric distance causes ICMEs to expand in size considerably (Démoulin and Dasso, 2009; Gulisano et al., 2010). By the time they reach 1 AU, ICMEs typically span around one-third of an AU in radial width (Jian et al., 2006; Klein and Burlaga, 1982) and several tens of degrees in heliocentric longitude (e.g. Good and Forsyth, 2016). In addition, the spherical geometry of the solar wind outflow gives rise to steep pressure gradients that may flatten ICMEs in the plane perpendicular to their propagation direction through a process known as pancaking (Riley and Crooker, 2004; Russell and Mulligan, 2002). A structured solar wind, in which solar wind speed varies with heliospheric latitude and longitude, can further distort an ICME's global shape (Owens, 2006; Savani et al., 2010). Distortions arising from solar wind interactions are more pronounced when the flow momentum exceeds the flux rope's magnetic tension force, which resists distortion. ICME morphology and propagation may also be significantly altered when ICMEs interact with each other (e.g. Lugaz et al., 2015; Lugaz et al., 2017). Magnetic reconnection can also erode ICME field structure by varying degrees in the corona and interplanetary space (e.g. Dasso et al., 2007; Ruffenach et al., 2015).

Given the various ways in which ICMEs can change during propagation, an interesting and open question is the extent to which the underlying flux rope structure of the ICME also changes with heliocentric distance. A related question is the extent to which ICME flux ropes evolve self-similarly, i.e., without significant changes or reorderings of field features. Understanding the nature of this radial evolution is important for understanding ICMEs as magnetohydrodynamic structures, and also for efforts to forecast the space weather impact of ICMEs (Lindsay et al., 1999; Kubicka et al., 2016; Möstl et al., 2018): would ICME flux ropes observed by an upstream space weather monitor (placed nearer to the Sun than the L1 point) have the same structure on arrival at the Earth? From the observational perspective, these questions may best be addressed by analyzing ICMEs observed by pairs of spacecraft located at the same or similar heliographic latitudes and longitudes (i.e., radially aligned) and separated in heliocentric distance. Such observations allow ICME evolution along fixed radial lines from the Sun to be determined. If the component of an ICME's propagation velocity perpendicular to the radial direction between the observing spacecraft is small, the same region of the ICME will be sampled at both spacecraft.

Instances of ICME encounters by aligned spacecraft have been rare and previous studies involving alignment observations have generally been restricted to case studies. As of 2015, ten events had been reported where the observing spacecraft were separated by less than 10° in heliographic latitude and longitude and by more than ~0.2 AU (Good et al., 2015, and references therein). These events were all observed before the launch of the Solar TErrestrial RElations Observatory (STEREO) mission. Burlaga et al. (1981) studied a magnetic cloud observed by Helios 1 (at 1AU) and the two Voyagers (both at 2 AU) while the spacecraft were separated by 10° in longitude in January 1978. A magnetic cloud observed by ACE at 1 AU and Ulysses at 5.4 AU has been extensively studied (Du et al., 2007; Li et al., 2017; Nakwacki et al., 2011; Skoug et al., 2000). Alignment observations of the Bastille Day event (14 July 2000) made by the ACE and NEAR spacecraft (at 1 and 1.8 AU, respectively) have also been analyzed and modeled in detail (Mulligan et al., 2001; Russell et al., 2003). One of the first and most significant studies to directly consider ICME magnetic field similarity at aligned spacecraft was performed by Mulligan et al. (1999). They analyzed four events observed by the Wind and NEAR spacecraft, three of which were observed when the spacecraft were separated by less than 12° in longitude and ~0.2-0.4 AU in radial distance; two of the events displayed similarities at the different spacecraft, while a third showed significant dissimilarities in field structure. Leitner et al. (2007) performed Lundquist fits (Lundquist, 1950; Burlaga, 1988; Lepping et al., 1990) to seven ICME flux ropes observed by multiple spacecraft during solar cycles 20 and 21; the spacecraft were located at heliocentric distances ranging from 0.62 to 9.4 AU, and were separated by up to 20° in heliographic longitude. Good et al. (2015) and Good et al. (2018) performed case studies of two more recently observed ICME flux ropes that displayed strong similarities at pairs of radially aligned spacecraft. Kubicka et al. (2016) used magnetic field observations of an ICME observed upstream of the Earth at 0.72 AU to accurately predict the Dst index subsequently observed at Earth. A notable study was performed by Winslow et al. (2016), who analyzed a complex event in which the field structure and orientation of an ICME showed significant differences at two lined-up spacecraft. Wang et al. (2018) have recently analyzed an ICME observed at four spacecraft during 2014, and found that the ICME's flux rope appeared to have been heavily eroded with a more twisted field near its core than in its outer layers. This finding challenges our current understanding of flux rope structure and formation processes in the solar atmosphere.

Recent planetary missions, namely the MErcury Surface, Space ENvironment, GEochemistry, and Ranging (MESSENGER) mission and Venus Express, combined with STEREO and the L1 spacecraft near 1 AU, have offered extensive in situ coverage of the inner heliosphere (i.e., up to ~1 AU in heliocentric distance from the Sun). During the time period when these missions were jointly active (~2006-15), multiple ICME flux ropes were observed by radially aligned spacecraft. In this work, we examine the magnetic field signatures of the 18 most prominent examples identified by Good and Forsyth (2016) from this time period. This work represents the first extensive investigation of ICME magnetic structure and its evolution that uses a relatively large set of radial alignment observations, and the first to use such a large dataset from the inner heliosphere. By studying a larger number of events, we have sought to determine trends in ICME properties that cannot be established with case studies.

Using a simple mapping technique (Good et al., 2018), the time series of the magnetic field in the flux ropes at the inner spacecraft have been mapped to the heliocentric distances of the outer spacecraft. The plots that result from this mapping allow for easy and direct comparison of the underlying field structure. Self-similarities (or, conversely, reorderings of field features) are often more discernible in these plots than in the unmapped data. The plots are also useful for confirming whether ICME signatures at different spacecraft are associated with the same ICME. For most of the ICMEs studied, there were significant similarities in field structure at the different observation points. Although the analysis primarily focuses on determining similarity in the time series profiles, Lundquist fitting has also been applied to characterize the global structure of the flux ropes at each spacecraft, and changes in the fitted parameters (e.g., rope axis orientation) between the aligned spacecraft are quantified and discussed.

In Section 2, the criteria used to identify the ICMEs are described and the ICME observations are presented. The time series mappings are presented in Section 3 and the Lundquist fits to the flux rope profiles are presented in Section 4. In Section 5, the results from the previous sections are considered in conjunction and discussed.

## 2 Data and Event List

### 2.1 Spacecraft Data

18 ICME flux ropes observed by pairs of spacecraft close to radial alignment have been selected for analysis. The ICMEs were originally identified by Good and Forsyth (2016) in their multipoint analysis of ICMEs encountered by MESSENGER and Venus Express, and a preliminary study of their properties was performed by Good (2016). The spacecraft line-ups were formed by pairings of MESSENGER (Solomon et al., 2001), Venus Express (Titov et al., 2006), Wind (Ogilvie and Desch, 1997), and the twin STEREOs (Kaiser, 2005).

Magnetic field data from magnetometers on board MESSENGER (MAG; Anderson et al., 2007), Venus Express (MAG; Zhang et al., 2008), STEREO (IMPACT MAG; Acuña et al., 2008), and Wind (MFI; Lepping et al., 1995) have been used. All field data used were at a 1 minute-averaged resolution and were obtained from the Heliospheric Cataloguing, Analysis and Techniques Service (HELCATS) project results. Solar wind plasma data from STEREO's PLASTIC instrument (Galvin et al., 2008) at 1 minute resolution and Wind's 3DP instrument (Lin et al., 1995) at ~24 second resolution have also been used. No continuous plasma data were available from MESSENGER or Venus Express while they were in the solar wind.

Magnetospheric intervals in the MESSENGER and Venus Express data have been removed. The intervals occurred two or three times every 24 hours in the MESSENGER dataset following the spacecraft's orbital insertion at Mercury in March 2011, and once every 24 hours throughout the Venus Express dataset. In cases where the ICME field rotations were not altered by the planetary bow shocks, the magnetosheath intervals have not been removed; although the ICME magnetic field rotations remained intact during such intervals, the field magnitudes were enhanced relative to the intrinsic fields of the ICMEs. The type of bow shock crossing (i.e., quasi-parallel or quasi-perpendicular) will determine whether the flux rope field direction is altered (Turc et al., 2014).

### 2.2 Event List

The 18 flux ropes were identified by (i) relatively smooth, monotonic rotations in the B-field direction coinciding with (ii) enhanced B-field magnitudes compared to the ambient solar wind that (iii) were observed for approximately 4 hours or more. Criteria (i) and (ii) are standard signatures of ICME flux ropes observed within the inner heliosphere (L. Burlaga et al., 1981), while criterion (iii) has been applied to exclude smaller flux ropes that may not be associated with ICMEs. Only B-field signatures have been used to identify the ICMEs. The flux rope leading and trailing edges were located at discontinuities in the magnetic field between which criteria (i) – (iii) were satisfied.

Table 1 lists the arrival times of the rope leading and trailing edges ($t_L$ and $t_T$, respectively) at each spacecraft, as well as the arrival times of any preceding discontinuities ($t_S$) in the magnetic field that bounded sheath regions. In some cases these discontinuities may be shocks, but formal shock identifications have not been made due to the absence of plasma data at Venus Express and MESSENGER. The absolute latitudinal and longitudinal separations of the spacecraft pairs ($\Delta\theta_{HCI}$ and $\Delta\varphi_{HCI}$, respectively) are also listed in Table 1, in heliocentric inertial coordinates.

Flux rope signatures at different spacecraft were judged to be associated with the same ICME if the arrival times were broadly consistent with typical and realistic propagation speeds. For example, an ICME observed at the orbital distance of Mercury would be expected to arrive at the orbit of Venus around 1-2 days later, and at 1 AU ~3 days later. Strict time windows were not imposed in order to allow for particularly fast or slow events. Only cases where the latitudinal and longitudinal separations of the observing spacecraft did not exceed 15° have been included in the analysis. The values of $\Delta\theta_{HCI}$ and $\Delta\varphi_{HCI}$ were generally less than 10°, with mean values of ~3° and ~5°, respectively. Although non-zero, the separations were small relative to the typical CME angular extent of ~50° to 60° (e.g., Yashiro et al., 2004) seen in coronagraph images, and small relative to reported ICME extents in interplanetary space (e.g., Good and Forsyth, 2016; Witasse et al., 2017). No requirement was placed on similarity of flux rope field structure when making the associations.

The 18 ICMEs have been classified according to the 'quality' of their signatures at the aligned spacecraft. Quality 1 (Q1) events tended to display relatively simple field rotations, smoother field magnitude profiles, longer durations, and unambiguous boundaries. A number of these events displayed field profiles consistent with magnetic cloud observations (e.g., Burlaga, 1988). Quality 3 (Q3) events, in contrast, tended to show more complex field rotations, shorter durations, less clearly defined boundaries, and stronger interactions with the ambient environment (i.e., the solar wind or other ICMEs). Quality 2 (Q2) events were intermediate cases. Classification of events in this way is somewhat subjective, but is nonetheless a useful exercise.

The B-field time series for the ICMEs are displayed according to Q-number in Figures 1 (Q1 events), 2 (Q2 events) and 3 (Q3 events). For each event, the time series at the inner spacecraft in the line-up is displayed in the upper panel, and the time series at the outer spacecraft in the panel beneath. For those events observed at Wind, STEREO-A or STEREO-B, the solar wind proton speed is also displayed. Flux rope boundaries are indicated by vertical dashed lines. The B-field data are displayed in Spacecraft Equatorial (SCEQ) coordinates, in which $z$ is parallel to the solar rotation axis, $y$ points to solar west, and $x$ completes the right-handed system. In the following subsections, the events in each of the three quality categories are briefly described.

### 2.2.1 Q1 Events

Observed during the deep minimum of Solar Cycle 24, Event 2 showed only a small field magnitude enhancement, and convected along with the solar wind without producing a sheath. The rotation and low variability of the B-field which characterize the flux rope are nonetheless clear. Events 8, 13, 14 and 18 displayed unmistakable ICME flux rope profiles with large central rises in the field magnitude. These four ICMEs were all preceded by sheaths. The development of a compression region (Fenrich and Luhmann, 1998; Kilpua et al., 2012) at the rear of Event 8 was evident, a result of the fast solar wind stream arriving at STEREO-B at the beginning of DoY 314. Event 16 has been classified as Q1 largely due to the quality of its signatures at Wind; the signatures at Venus Express were less clear, with a large magnetospheric cut-out near the flux rope midpoint that was followed by a magnetosheath interval in which the field strength rose significantly. Event 8 has previously been studied by Good and Forsyth (2016), Good et al. (2018) and Amerstorfer et al., (2018), Event 14 by Good et al. (2015), and Event 16 by Kubicka et al. (2016) and Palmerio et al. (2018).

### 2.2.2 Q2 Events

As in the case of Q1 Event 8, Events 3, 5 and 6 were observed during solar minimum, and also showed relatively small B-field enhancements without shocks or sheaths. Events 3 and 5 were both embedded in regions of steadily declining solar wind proton speed. Event 3 was of short-duration, but clearly showed the B-field signatures of a flux rope. A significant compression region had developed at the rear of Event 6 by the time it arrived at STEREO-A, presumably due to a trailing fast stream. Events 10 and 11 were both moderately fast ICMEs that produced prominent shocks and sheaths. The shock driven by Event 10 had propagated into a slower-moving structure ahead, possibly another ICME, by the time of observation at STEREO-A. There was only a short period of field rotation in Event 15 behind the ICME leading edge at both spacecraft, with the remainder of the field being close to radial; this ICME has been previously studied by Rollett et al. (2014).

### 2.2.3 Q3 Events

Events 1, 4, and 7 were observed during solar minimum, with Events 1 and 7 displaying relatively low field magnitudes. Events 4 and 7 were embedded in slow solar wind, and both produced extended sheaths with weak leading shocks at STEREO-B. The in situ signatures of Event 4 at STEREO-B have previously been studied by Möstl et al. (2011). The relatively high-magnitude field seen to the rear of Event 4 at the inner spacecraft was not observed at the outer. A shock was present near the midpoint of Event 7 at STEREO-B, driven by an overtaking ICME; the field magnitude and speed profile of this short-duration event were perturbed significantly by the shock wave. Event 9, also of short duration, displayed a prominent sheath region. By the time of arrival at Venus Express, Event 12 was strongly interacting with a preceding ICME; the ICMEs were observed separately at MESSENGER, prior to the onset of the interaction. Event 17 displayed complex field component variations and a rise in the field magnitude from front to back that was possibly due to a trailing fast stream.

We note that the durations of Events 1, 7 and 17 were greater at the inner spacecraft than at the outer spacecraft, in contrast to the other ICMEs. In the case of Event 1, we suggest that this was due to the spacecraft having sampled significantly different regions of the ICME; although the spacecraft longitudinal separation (at 13°) was within the selection limit of 15°, it was the largest of the 18 alignments analyzed. Compression arising from external interactions may have contributed to the shorter outer-spacecraft durations observed for Events 7 and 17.

| Event # | s/c 1 | $R_1$ [AU] | s/c 2 | $R_2$ [AU] | $\Delta\theta_{HCI}$ [°] | $\Delta\Phi_{HCI}$ [°] | s/c 1 | | | | | | s/c 2 | | | | | | Q |
|---|---|---|---|---|---|---|---|---|---|---|---|---|---|---|---|---|---|---|---|
| | | | | | | | $t_S$ | | $t_L$ | | $t_T$ | | $t_S$ | | $t_L$ | | $t_T$ | | |
| 1 | MES | 0.575 | VEX | 0.719 | 5.1 | 13.5 | - | - | 2007 May 4 | 20:57 | 2007 May 6 | 00:58 | - | - | 2007 May 6 | 05:38 | 2007 May 6 | 20:53 | 3 |
| 2 | VEX | 0.723 | STB | 1.026 | 1.1 | 8.8 | - | - | 2008 Dec 29 | 20:46 | 2008 Dec 30 | 10:22 | - | - | 2008 Dec 31 | 03:39 | 2009 Jan 1 | 00:53 | 1 |
| 3 | VEX | 0.721 | STB | 1.015 | 0.7 | 1.8 | - | - | 2009 Jan 17 | 14:36 | 2009 Jan 17 | 22:49 | - | - | 2009 Jan 19 | 01:48 | 2009 Jan 19 | 15:03 | 2 |
| 4 | MES | 0.326 | STB | 1.002 | 5.4 | 1.0 | - | - | 2009 Feb 15 | 04:24 | 2009 Feb 15 | 14:12 | - | - | 2009 Feb 18 | 18:42 | 2009 Feb 19 | 11:15 | 3 |
| 5 | VEX | 0.728 | STA | 0.956 | 3.1 | 9.7 | 2009 Jun 2 | 16:11 | 2009 Jun 2 | 18:39 | 2009 Jun 3 | 20:12 | - | - | 2009 Jun 3 | 03:45 | 2009 Jun 4 | 21:26 | 2 |
| 6 | VEX | 0.727 | STA | 0.957 | 3.5 | 10.1 | - | - | 2009 Jul 10 | 10:38 | 2009 Jul 11 | 06:24 | - | - | 2009 Jul 11 | 23:07 | 2009 Jul 13 | 05:37 | 2 |
| 7 | MES | 0.561 | STB | 1.084 | 5.5 | 3.6 | - | - | 2009 Aug 28 | 01:22 | 2009 Aug 28 | 15:36 | 2009 Aug 30 | 02:48 | 2009 Aug 30 | 19:49 | 2009 Aug 31 | 08:25 | 3 |
| 8 | MES | 0.465 | STB | 1.083 | 7.0 | 1.3 | 2010 Nov 5 | 11:46 | 2010 Nov 5 | 16:52 | 2010 Nov 6 | 13:08 | 2010 Nov 7 | 19:05 | 2010 Nov 8 | 03:24 | 2010 Nov 9 | 09:04 | 1 |
| 9 | VEX | 0.727 | STA | 0.958 | 1.0 | 4.6 | 2011 Mar 18 | 20:46 | 2011 Mar 19 | 04:54 | 2011 Mar 19 | 11:54 | 2011 Mar 19 | 11:24 | 2011 Mar 19 | 23:36 | 2011 Mar 20 | 10:18 | 3 |
| 10 | VEX | 0.727 | STA | 0.957 | 1.2 | 2.5 | 2011 Mar 22 | 08:51 | 2011 Mar 22 | 17:28 | 2011 Mar 23 | 18:20 | 2011 Mar 22 | 18:20 | 2011 Mar 23 | 06:56 | 2011 Mar 25 | 01:05 | 2 |
| 11 | VEX | 0.728 | STA | 0.957 | 1.5 | 5.0 | 2011 Apr 5 | 07:08 | 2011 Apr 5 | 17:12 | 2011 Apr 6 | 15:03 | 2011 Apr 5 | 21:43 | 2011 Apr 6 | 09:40 | 2011 Apr 7 | 20:28 | 2 |
| 12 | MES | 0.322 | VEX | 0.725 | 0.2 | 6.6 | - | - | 2011 Jun 5 | 04:31 | 2011 Jun 5 | 09:37 | - | - | 2011 Jun 5 | 13:19 | 2011 Jun 5 | 23:28 | 3 |
| 13 | MES | 0.460 | VEX | 0.725 | 2.6 | 2.0 | 2011 Oct 15 | 08:25 | 2011 Oct 15 | 11:17 | 2011 Oct 16 | 06:23 | 2011 Oct 16 | 00:50 | 2011 Oct 16 | 06:58 | 2011 Oct 17 | 09:39 | 1 |
| 14 | MES | 0.439 | STB | 1.086 | 6.8 | 4.3 | 2011 Nov 4 | 15:09 | 2011 Nov 5 | 00:43 | 2011 Nov 5 | 17:05 | 2011 Nov 6 | 05:10 | 2011 Nov 6 | 22:57 | 2011 Nov 8 | 17:48 | 1 |
| 15 | MES | 0.316 | VEX | 0.719 | 4.4 | 2.8 | 2012 Mar 7 | 05:04 | 2012 Mar 7 | 06:10 | 2012 Mar 7 | 16:28 | 2012 Mar 7 | 13:26 | 2012 Mar 7 | 20:14 | 2012 Mar 8 | 11:43 | 2 |
| 16 | VEX | 0.727 | Wind | 1.006 | 0.2 | 5.4 | - | - | 2012 Jun 15 | 19:26 | 2012 Jun 16 | 08:28 | 2012 Jun 16 | 19:35 | 2012 Jun 16 | 22:15 | 2012 Jun 17 | 11:43 | 1 |
| 17 | MES | 0.420 | VEX | 0.722 | 0.1 | 7.3 | - | - | 2012 Dec 15 | 21:24 | 2012 Dec 16 | 19:52 | - | - | 2012 Dec 17 | 12:19 | 2012 Dec 18 | 05:02 | 3 |
| 18 | VEX | 0.725 | STA | 0.960 | 1.4 | 5.5 | 2013 Jan 8 | 09:22 | 2013 Jan 8 | 15:24 | 2013 Jan 9 | 19:48 | 2013 Jan 9 | 02:25 | 2013 Jan 9 | 10:39 | 2013 Jan 10 | 17:17 | 1 |

**Table 1.** Details of the spacecraft alignments and ICME arrival times. 'MESSENGER' is abbreviated to 'MES', 'Venus Express' to 'VEX', 'STEREO-A' to 'STA', and 'STEREO-B' to 'STB'. The heliocentric distances of the inner and outer spacecraft in each pair ($R_1$ and $R_2$, respectively) and the spacecraft's latitudinal and longitudinal separations in HCI coordinates ($\Delta\theta_{HCI}$ and $\Delta\varphi_{HCI}$, respectively) are given. Also listed are the arrival times of the preceding shock ($t_S$), flux rope leading edge ($t_L$) and flux rope trailing edge ($t_T$) for each ICME at each spacecraft in the observing spacecraft line-up, and the quality classification of the observations (Q).

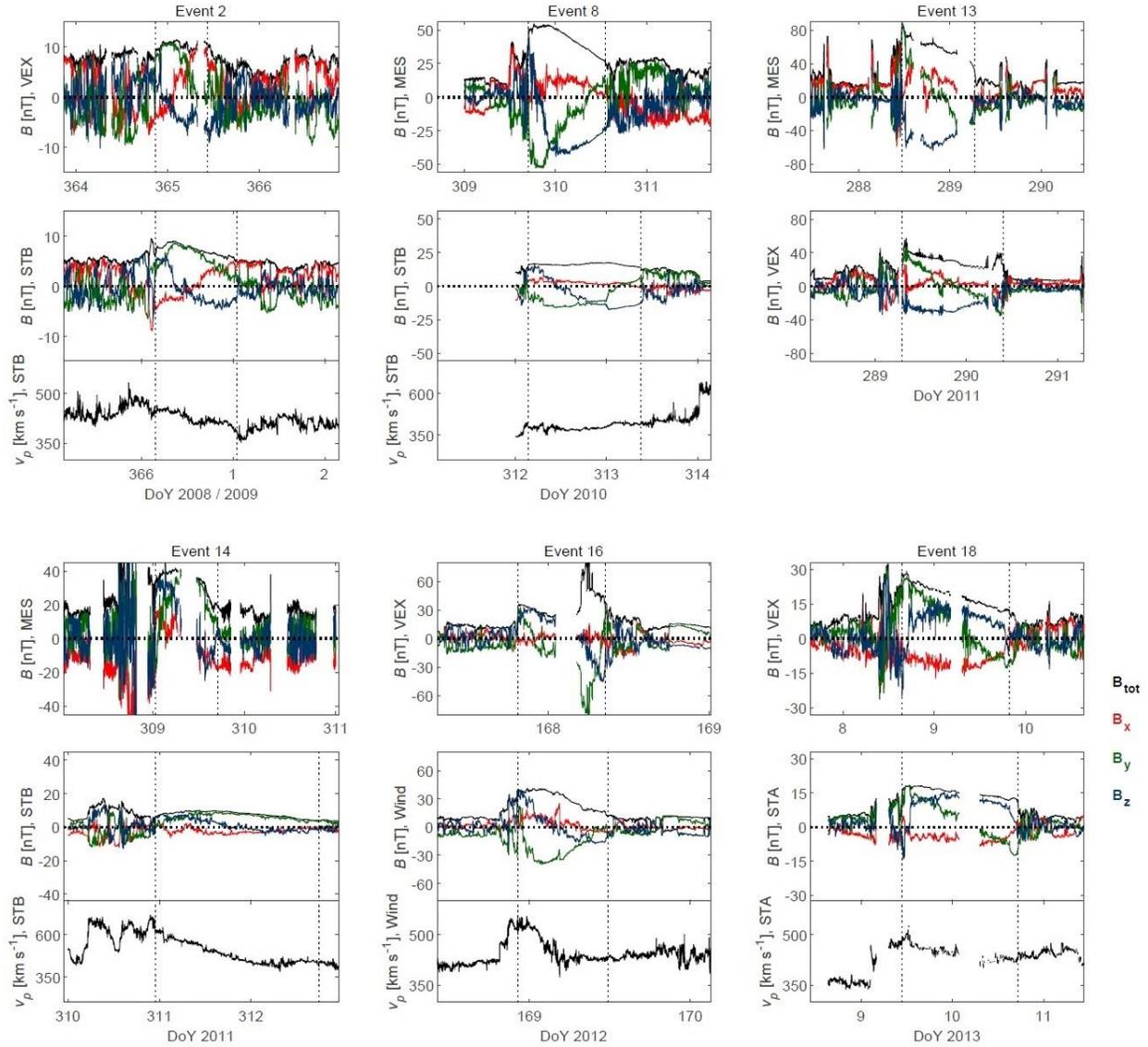

**Figure 1.** Quality 1 ICME observations. B-field data in SCEQ coordinates at the inner spacecraft are shown in the upper panel for each event subplot, and at the outer spacecraft in the panel below. Where the outer spacecraft was Wind, STEREO-A or STEREO-B, the solar wind proton speed is also displayed. Flux rope boundaries are indicated with vertical dotted lines. For ease of comparison, the *y*-axis scaling for the B-field data and the *x*-axis timespan is the same at both spacecraft for each event.

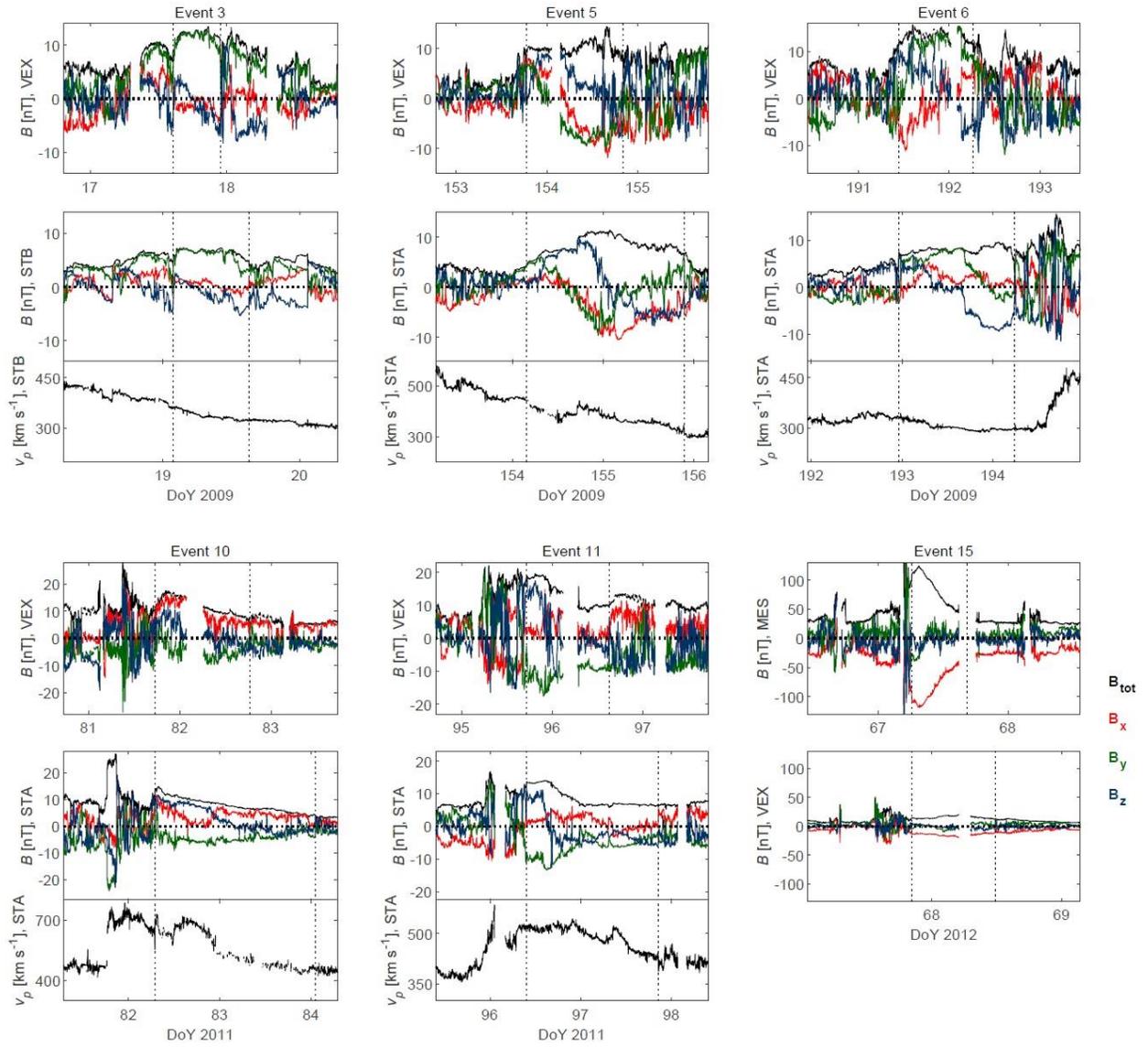

**Figure 2.** Quality 2 ICME observations. (Same format as Figure 1.)

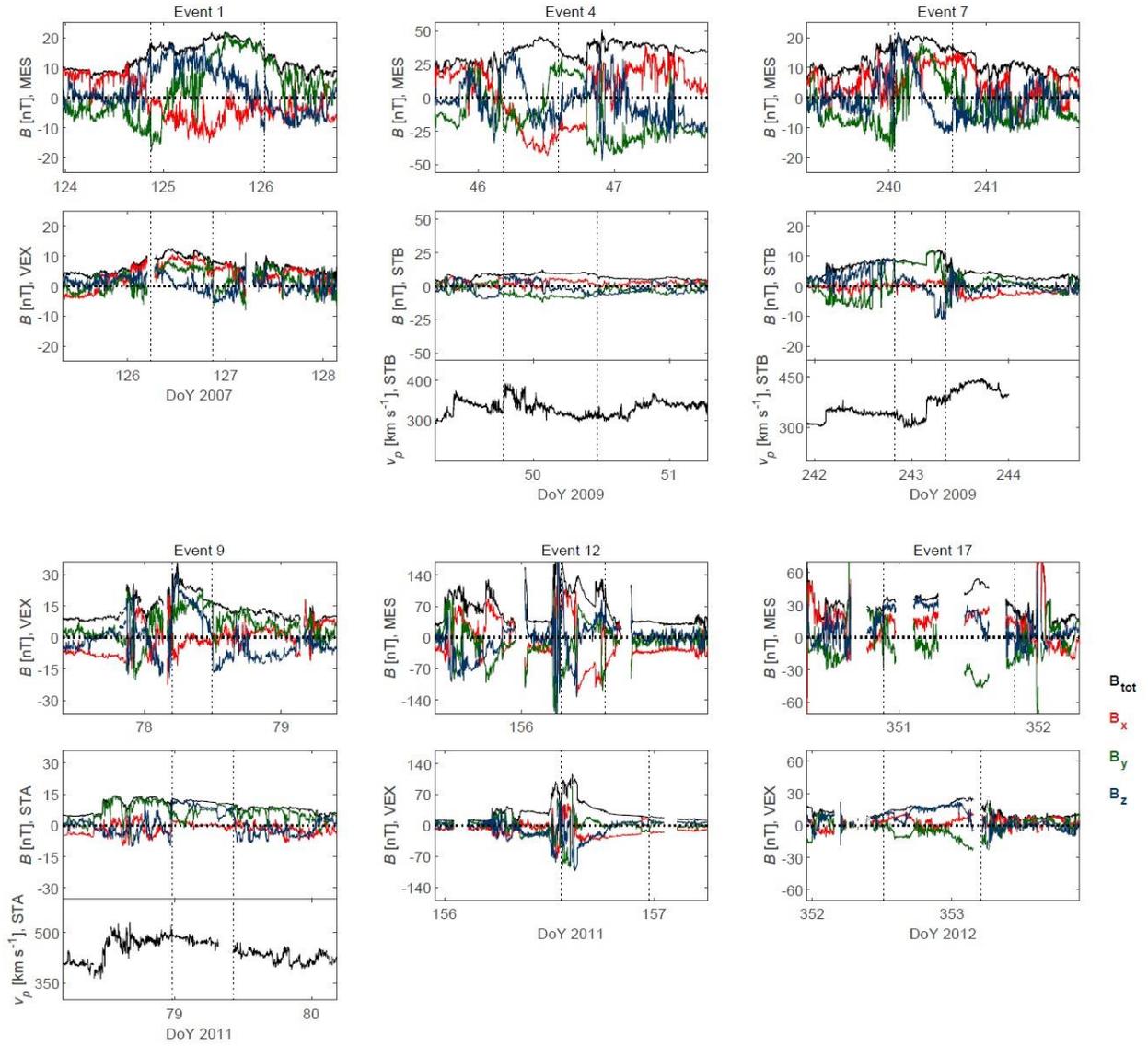

**Figure 3.** Quality 3 ICME observations. (Same format as Figure 1.)

## 3 Profile mapping

### 3.1 Mapping technique

For each ICME, we now compare the underlying flux rope profile structure at the inner spacecraft to the structure at the outer spacecraft. To facilitate this comparison, the empirical mapping technique described by Good et al. (2018), in which the inner-spacecraft B-field profiles are mapped to the heliocentric distances of the outer spacecraft, has been applied to the ICME time series. The technique imagines each B-field vector measured within the flux rope at the inner spacecraft being frozen-in to a discrete, radially propagating plasma parcel, and involves estimating the arrival time of each parcel at the outer-spacecraft distance. The mean speeds of the flux rope leading and trailing edges during their propagation between the spacecraft are determined from their arrival times, and a linear, mean speed profile across the rope is determined from these two speeds. The mean speed profile is then used to map each vector-parcel from the inner spacecraft to the outer-spacecraft distance. For an ICME expanding in the radial direction, the mapping effectively stretches the inner-spacecraft rope profile to the duration of the rope at the outer spacecraft.

Figures 4, 5 and 6 show the mappings for the Q1, Q2 and Q3 events, respectively. The panels for each event mapping show (from top to bottom) the $\hat{B}_x$, $\hat{B}_y$ and $\hat{B}_z$ field components, the latitude angle of the field direction, $\theta_B$, and the longitude angle, $\varphi_B$. Dark-colored lines in the panels show the profiles at the outer spacecraft, and pale-colored lines show the mapped flux rope profiles from the inner spacecraft; the inner-spacecraft rope vectors are plotted versus their estimated times of arrival at the heliocentric distance of the outer spacecraft. The mapping constrains the leading and trailing edge vectors of the inner and outer-spacecraft profiles to overlap. Flux rope boundaries are marked by vertical dashed lines in the figures. The vectors have been normalized to their magnitudes in order to remove any differences within the ropes that are due to changes in field strength, allowing easier comparison of the underlying rope structure.

Similarity in the profiles is partly assessed with two measures, namely the root-mean square error, $\epsilon$, and the mean absolute error, $\mu$, given by

$$\epsilon = \sqrt{\sum_{i=1}^{N}(B_{i,2} - B_{i,1})^2/N} \qquad (1)$$

$$\mu = \sum_{i=1}^{N}|B_{i,2} - B_{i,1}|/N \qquad (2)$$

respectively, where $N$ is the total number of data points in the time series, $i = 1, \ldots, N$, $B_{i,2}$ is the $i$th value of the field component at the outer spacecraft, and $B_{i,1}$ is the corresponding value in the mapping from the inner spacecraft. Values of $\epsilon$ and $\mu$ are calculated for each component in each mapping shown in Figures 4, 5 and 6. Equation 2 gives the mean value of $\mu_i = |B_{i,2} - B_{i,1}|$ for each component, where there are $N$ values of $\mu_i$ for each component. The standard deviations of $\mu_i$ are given as uncertainties. $\epsilon$ and $\mu$ range between 0 and 2 in value; lower values of $\epsilon$ and $\mu$ suggest a greater degree of similarity. Since they are functions of normalized vector components, $\epsilon$ and $\mu$ are unitless quantities. Similarity is also assessed through qualitative visual inspection of the overlap plots.

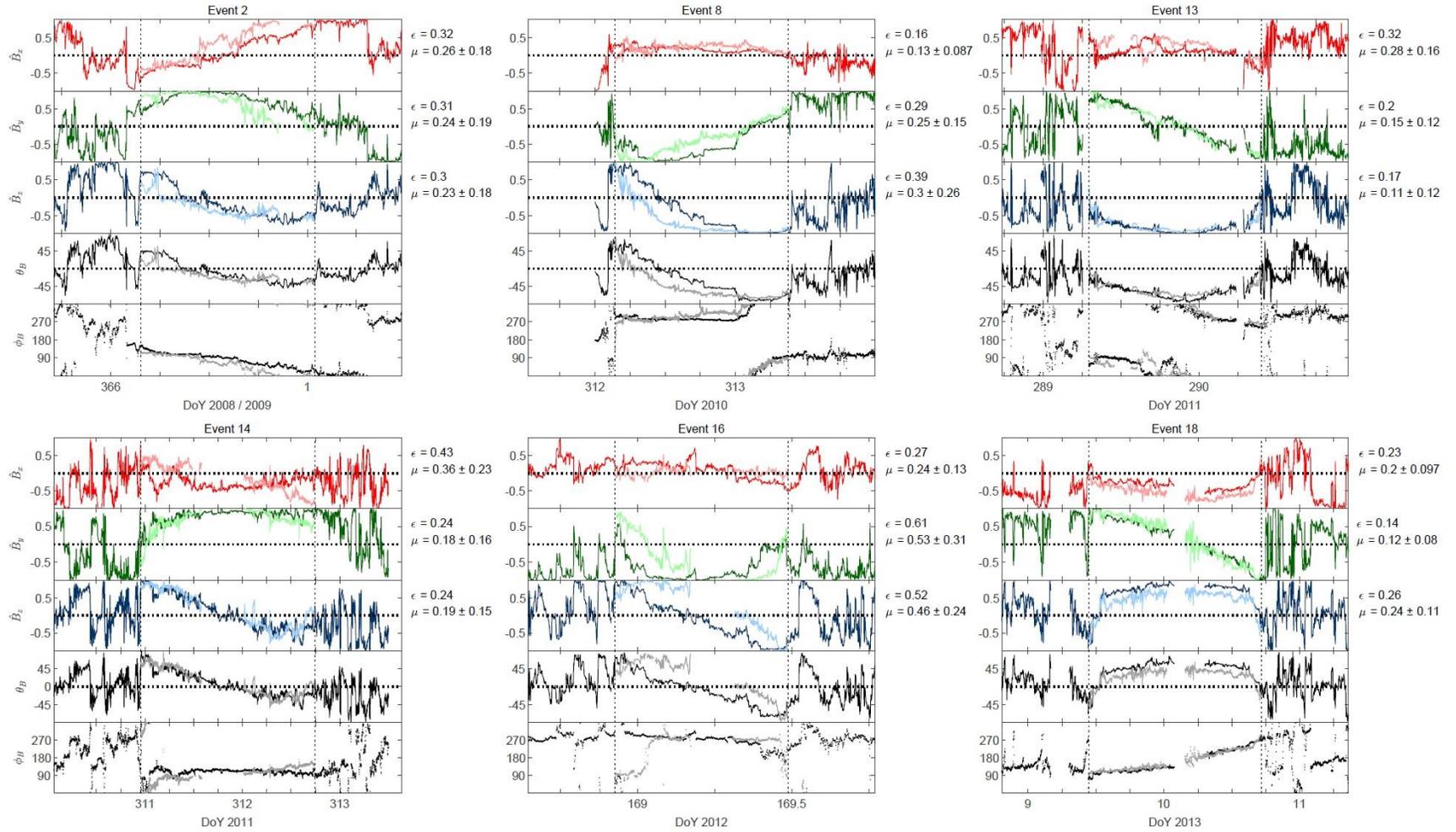

**Figure 4.** Mappings of the Quality 1 ICME profiles, in which the flux rope intervals at the inner spacecraft have been mapped to the heliocentric distances of the outer spacecraft. Pale-colored lines show the mapped profiles overlaying the rope intervals at the outer spacecraft, shown by the dark-colored lines. Flux rope boundaries are indicated with vertical dotted lines.

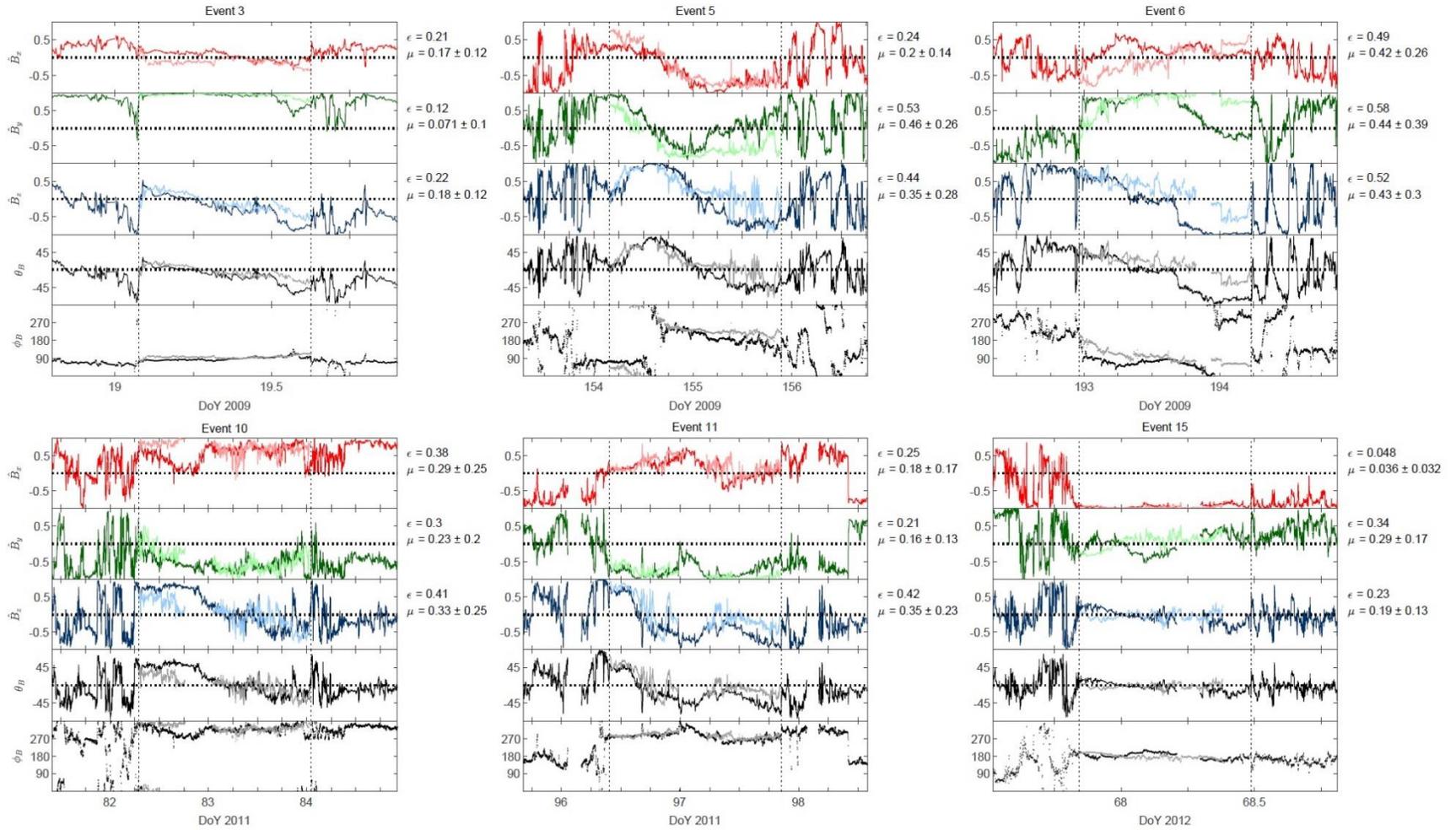

**Figure 5.** Mappings of the Quality 2 ICME profiles. (Same format as Figure 4.)

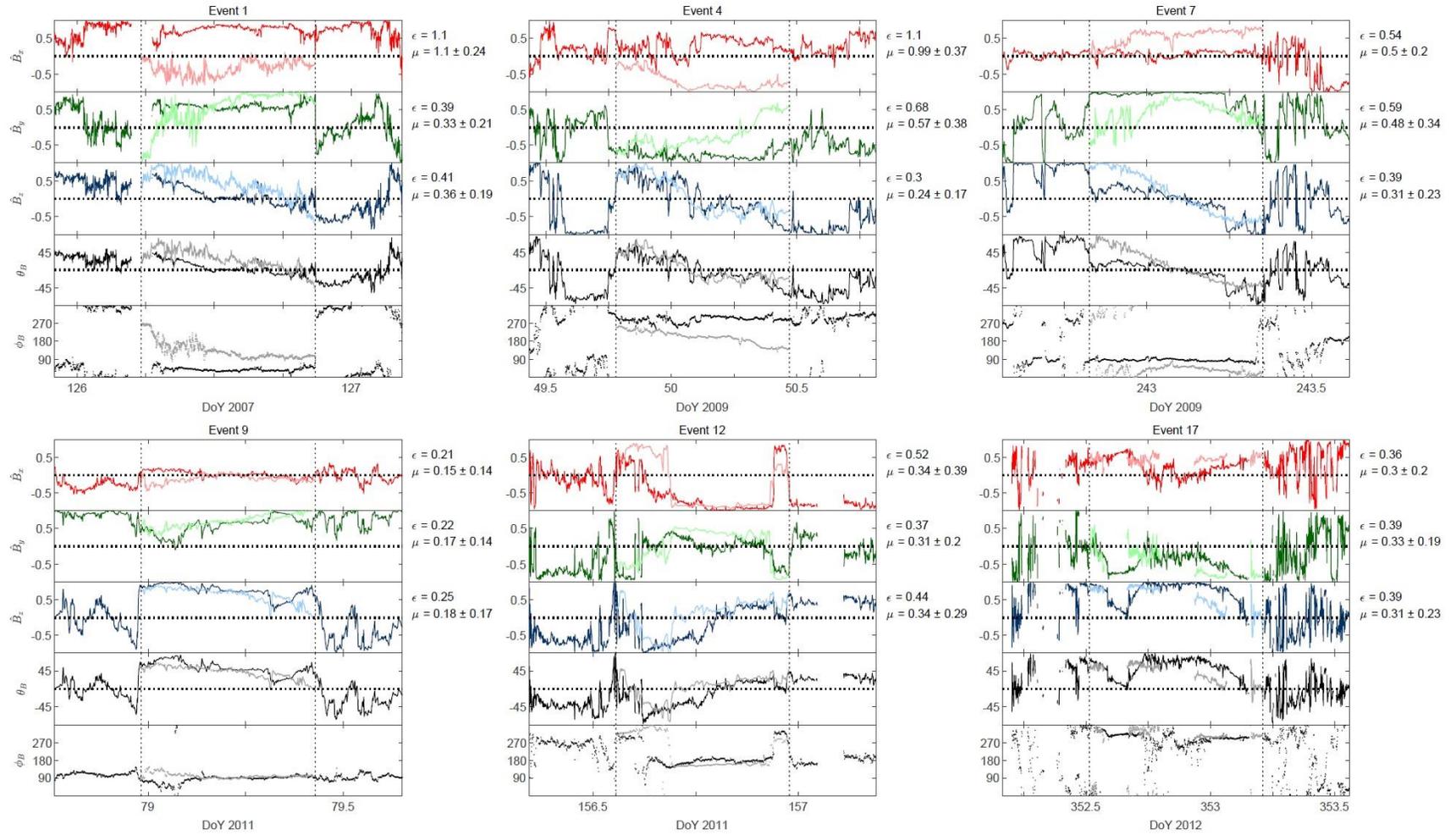

**Figure 6.** Mappings of the Quality 3 ICME profiles. (Same format as Figure 4.)

| | $\bar{\epsilon}$ | | | $\bar{\mu}$ | | |
|----|----|----|----|----|----|----|
| | $x$ | $y$ | $z$ | $x$ | $y$ | $z$ |
| Q1 | $0.29 \pm 0.09$ | $0.30 \pm 0.16$ | $0.31 \pm 0.12$ | $0.25 \pm 0.08$ | $0.26 \pm 0.15$ | $0.26 \pm 0.12$ |
| Q2 | $0.27 \pm 0.15$ | $0.35 \pm 0.18$ | $0.37 \pm 0.12$ | $0.22 \pm 0.13$ | $0.28 \pm 0.15$ | $0.31 \pm 0.10$ |
| Q3 | $0.64 \pm 0.38$ | $0.44 \pm 0.17$ | $0.36 \pm 0.07$ | $0.56 \pm 0.39$ | $0.37 \pm 0.14$ | $0.29 \pm 0.07$ |

**Table 2.** Mean values of the root-mean square error, $\epsilon$, and the mean absolute error, $\mu$, for each field component mapping in each event quality category. The standard deviations are given as uncertainty values.

## 3.2 Profile similarity

Significant macroscale similarities in the field component profiles can be seen across all six Q1 event mappings in Figure 4. The sense of rotation of the magnetic field vector as indicated by the $\theta_B$ and $\varphi_B$ profiles is consistent in all mappings. The mean values and standard deviations of $\epsilon$ and $\mu$ listed in Table 2 for the Q1 events are all relatively low. There is less similarity in the Event 16 mapping compared to the others.

As with the Q1 events, there are significant macroscale similarities in the Q2 event profiles shown in Figure 5. Strong similarities are seen even in cases where the component profiles and field rotations are complex (e.g., Event 11). There is markedly less similarity in the Event 6 mapping compared to the others, although the sense of rotation in the mapping is broadly consistent. The mean values of $\epsilon$ and $\mu$ listed in Table 2 for the Q2 events are comparable to the corresponding Q1 values, although there is more variation in the values across the components.

The mappings for Q3 Events 9, 12 and 17 display broadly similar macroscale features in the field components, while the Q3 Event 1, 4 and 7 mappings show some significant dissimilarities that are reflected by the relatively large values of $\epsilon$ and $\mu$; there is a particularly low degree of overlap in the $x$ and $y$ components for these latter three events.

## 3.3 ICME kinematics

We now briefly consider some of the kinematic properties of the ICMEs analyzed. The mapping procedure described in the previous section involved finding the mean radial transit speeds of the flux rope leading and trailing edges, $v_L^M$ and $v_T^M$, respectively. These speeds are derived simply from the radial separation distances and transit times between the spacecraft. The mean center-of-mass transit speed is defined as $v_C^M = (v_L^M + v_T^M)/2$, and the mean expansion speed during transit is defined as $v_{EXP}^M = v_L^M - v_T^M$.

The spacecraft located near 1 AU (STEREO-A, STEREO-B and Wind) all made in situ measurements of the solar wind proton speed. For 12 alignments that included one of these three spacecraft, we compare the mean $v_L^M$, $v_T^M$, $v_C^M$ and $v_{EXP}^M$ speeds to their instantaneous values measured at 1 AU. In all cases, the 1 AU spacecraft were further from the Sun in the aligned spacecraft pair. The mean speeds and 1 AU speeds for the 12 events are listed in Table 3. In situ values of $v_L$, $v_T$ and $v_C$ are taken as the 15-minute averages of the radial proton speed immediately after the leading edge observation time, immediately before the trailing edge observation time, and centered on the rope time series midpoint, respectively. The 1 AU expansion speed, $v_{EXP}$, is given by $v_L - v_T$. Note that Event 7 has not been included in this analysis because its speed profile had been strongly perturbed by interaction with another ICME by the time of arrival at 1 AU.

Figure 7 shows the mean characteristic speeds plotted versus the corresponding instantaneous values at 1 AU. The dashed lines in the figure are $x = y$ lines. Points above the line indicate deceleration (i.e., the mean transit speed exceeded the instantaneous speed at the outer spacecraft) and points below the line indicate acceleration (i.e., the instantaneous speed at the outer spacecraft exceeded the mean transit speed), and points on the line indicate constant speed.

We also consider whether $v_L^M$, $v_T^M$ and $v_C^M$ were correlated with the speed of the ambient solar wind. For each ICME observed at 1 AU, 6-hour averages of the solar wind proton speed were taken directly ahead of any interaction (i.e., sheath) region at the front of the ICME and directly following any interaction region to the rear of the ICME. The interaction regions were identified by enhanced magnetic field magnitudes and increased field component variability relative to the ambient solar wind. We assume that the solar wind speed propagated at constant speed, so that the speeds measured at 1 AU will have been the same as those encountered by the ICME throughout its transit between the spacecraft. In panel (a) and (b) of Figure 7, black (red) data points indicate cases where the solar wind speed ahead of the ICME were less (greater) than $v_L^M$ and $v_C^M$, respectively. In panel (c), which shows $v_T^M$, the comparison is made to the solar wind speed to the rear of the ICME.

In Figure 7, panel (a) it can be seen that the mean transit speeds of the rope leading edges were generally very similar to their speeds at 1 AU, with some spread in values about the constant-speed line. In 9 of 12 cases, $v_L^M$ exceeded the solar wind speed ahead of the ICME. The center-of-mass speeds showed a similar pattern to the leading edge speeds.

In panel (c), it is notable that the trailing edge speed at 1 AU exceeded $v_T^M$ (implying acceleration) in 10 of 12 cases, in contrast to the leading edge and center-of-mass behavior. In 9 of the 10 cases where $v_T^M$ was less than $v_T$, the solar wind speed to the rear of the ICME exceeded $v_T^M$; the increase in speed of the trailing edge up to 1 AU may have been driven by interaction with the faster solar wind in these cases.

Panel (d) shows that $v_{EXP}^M$ exceeded $v_{EXP}$ at 1 AU in most cases, indicating that expansion speeds reduced with propagation to 1 AU. Reductions in $v_{EXP}$ were caused primarily by increases in the trailing edge speed rather decreases of the leading edge speed. The expansion speed remained approximately constant in four cases, and one case displayed an apparent increase in $v_{EXP}$ up to 1 AU.

We note that Event 5 displayed anomalously high mean leading edge and center-of-mass speeds that were not consistent with the speeds measured at 1 AU. In this case, the assumption that both spacecraft observed approximately the same region of the ICME may not have been valid, and the spacecraft angular separation may have been significant. For example, the ICME may have been propagating in solar wind with a highly sheared speed profile, with the ICME front along the radial line to STEREO-A embedded in faster solar wind and running ahead of the front observed by Venus Express. The ICME observed at STEREO-A was indeed propagating with relatively fast ambient solar wind; it cannot be determined whether the ICME at Venus Express was embedded in slower wind. Owens et al. (2017) consider the implications of ICMEs propagating in solar wind with a highly sheared speed profile.

| Event # | $v_L{}^M$ [km s$^{-1}$] | $v_T{}^M$ [km s$^{-1}$] | $v_C{}^M$ [km s$^{-1}$] | $v_{EXP}{}^M$ [km s$^{-1}$] | $v_L$ [km s$^{-1}$] | $v_T$ [km s$^{-1}$] | $v_C$ [km s$^{-1}$] | $v_{EXP}$ [km s$^{-1}$] | $v_{SW}$ ahead [km s$^{-1}$] | $v_{SW}$ behind [km s$^{-1}$] |
|---|---|---|---|---|---|---|---|---|---|---|
| 2 | 408 | 327 | 367 | 81 | 442 | 390 | 413 | 52 | 467 | 382 |
| 3 | 347 | 304 | 325 | 43 | 360 | 318 | - | 41 | 379 | 321 |
| 4 | 325 | 301 | 313 | 24 | 342 | 305 | 333 | 36 | 286 | 310 |
| 5 | 1042 | 376 | 709 | 666 | 441 | 330 | 378 | 112 | 449 | 303 |
| 6 | 262 | 202 | 232 | 60 | 328 | 299 | 301 | 29 | 320 | 537 |
| 8 | 437 | 380 | 408 | 57 | 396 | 422 | 393 | -25 | 294 | 609 |
| 9 | 513 | 429 | 471 | 85 | 492 | 442 | 471 | 50 | 402 | 439 |
| 10 | 710 | 311 | 510 | 399 | 681 | 457 | 508 | 224 | 465 | 449 |
| 11 | 578 | 323 | 451 | 255 | 522 | 440 | 484 | 82 | 386 | 423 |
| 14 | 580 | 370 | 475 | 210 | 619 | 422 | 472 | 197 | 477 | 427 |
| 16 | 432 | 426 | 429 | 7 | 524 | 429 | 433 | 94 | 416 | 434 |
| 18 | 507 | 455 | 481 | 53 | 482 | 427 | 445 | 55 | 354 | 426 |

**Table 3.** Mean radial transit speeds of the flux rope leading edges ($v_L^M$), trailing edges ($v_T^M$), centers of mass ($v_C^M$), and the mean expansion speeds ($v_{EXP}^M$). The corresponding instantaneous speeds measured at 1 AU ($v_L$, $v_T$, $v_C$ and $v_{EXP}$) are also listed, as well as the solar wind speeds ($v_{SW}$) measured ahead of and behind the ICMEs at 1 AU.

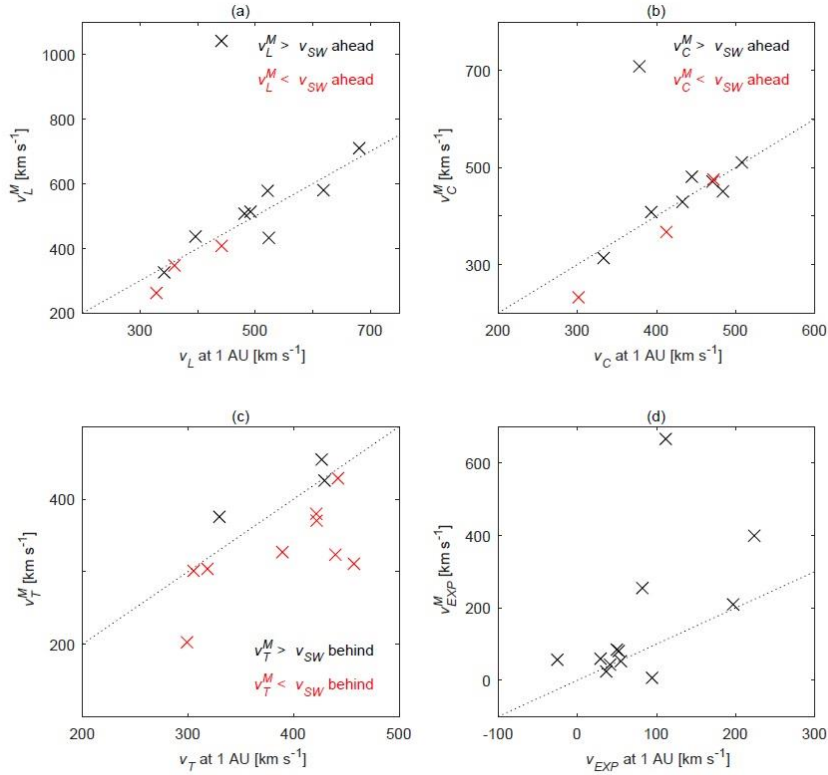

**Figure 7.** Plots of characteristic speeds at 1 AU (*x* axes) versus the corresponding mean speeds (*y* axes) during transit between the spacecraft. The panels show (a) the rope leading edge speeds, (b) the center-of-mass speeds, (c) the trailing edge speeds, and (d) the expansion speeds. Dotted lines are $x = y$ lines. Data-point color indicates whether the mean speeds are greater or less than the preceding solar wind speed at 1 AU in panels (a) and (b), and greater or less than the trailing solar wind speed at 1 AU in panel (c).

## 4 Lundquist fitting

### 4.1 Fitting model

The flux ropes have been fitted using the linear force-free Lundquist solutions. Such fitting allows global parameters of a flux rope to be estimated from the local measurements taken along the spacecraft trajectory through the rope. The Lundquist solutions were first used to model interplanetary flux ropes by Burlaga (1988) and Lepping et al. (1990). The simplest form of the Lundquist solutions models flux ropes as static, locally straight cylindrical tubes, in which the axial, tangential and radial B-field components relative to the cylinder axis are given by

$$B_A = B_0 J_0(\alpha R) \qquad (3a)$$
$$B_T = H B_0 J_1(\alpha R) \qquad (3b)$$
$$B_R = 0 \qquad (3c)$$

respectively, where $H$ is the rope handedness, $J_0$ and $J_1$ are the zeroth and first order Bessel functions, respectively, $B_0$ is the field strength along the axis, $R$ is the radial distance from the axis, and $\alpha$ is a constant. In order to convert the model spatial profiles to model time series that can be fitted to data, a linear relationship between spatial and temporal coordinates has been assumed, equivalent to assuming a uniform speed profile along the spacecraft trajectory. This simplifying assumption has been made to give consistency in the fitting across all spacecraft datasets. Although there are other fitting methods that account for non-uniform speed profiles (e.g., Farrugia et al., 1995), the speed profiles within the flux ropes observed by MESSENGER and Venus Express were not known. The boundaries of the model flux rope have been located at the first zero of $J_0$, at which $B_A = 0$. The model spatial units have been normalized by setting $\alpha = 1$ so that the rope radius $R_0 = 2.405$ in model units for all of the fits: these $\alpha$ and $R_0$ values give $J_0(\alpha R_0) = 0$ at the rope boundary.

A least-squares fitting procedure has been applied. The procedure involves the minimization of a chi-square parameter,

$$\chi^2 = \sum_{i=1}^{N} \sum_j (B_{ij}^O - B_{ij}^M)^2 / N \qquad (4)$$

where superscripts $O$ and $M$ refer to observed and model components, respectively, $N$ is the number of vectors in the fitted flux rope, $i = 1, \dots, N$, and $j = x, y, z$ denotes the Cartesian field components. The flux rope orientation is determined in the first stage of the fitting. This first stage is performed with normalized field data, which produces fits that better capture the field direction variations in the ropes. Minimum variance analysis (Goldstein, 1983) is used to give an initial estimate of the rope's orientation, and $\chi^2$ is then iteratively minimized using the Nelder-Mead simplex method to obtain a convergent fit (F1). The minimization process is then repeated with F1 as the initial estimated orientation to obtain a second convergent fit (F2), and repeated once again with F2 as the initial estimate to give a third convergent fit (F3). F2 is taken as the final fit and the angle between the orientation of F2 and F3, $\delta$, is used as a fit robustness measure. The successive fittings have been applied to reduce sensitivity to the initial estimated orientation; differences in the three fits arise when there are different, nearby local minima in the $\chi^2$ function. In the second stage of fitting, $B_0$ is obtained analytically from a one-parameter chi-square minimization (Lepping et al., 2003) using the un-normalized magnetic field data.

Key parameters of the fits are listed in Table 4. These include $H$, $B_0$, the latitude angles, $\theta_0$, and longitude angles, $\varphi_0$, of the rope's central axis direction relative to the solar equatorial plane, the $y_0$ parameter, which gives the distance in the $x$-$y$ plane between the spacecraft and rope axis at closest approach, the spacecraft impact parameter $p$, which ranges in value from 0 (where the spacecraft trajectory intersects the rope axis) to 1 (where the closest-approach distance to the axis is equal to the rope's cross-sectional radius), and the $D$ coefficient, which relates the observed width of the rope in the anti-sunward $x$ direction, $S$, to the fitted rope cylinder diameter, $S_R$, such that $S_R = DS$. Positive (negative) $y_0$ values indicate that the flux rope axis intersected the solar equatorial plane to the west (east) of the spacecraft location. The $y_0$ values have been normalized to the rope radius. The model geometry is shown in Figure 2 of the work by Burlaga (1988). For each event, a fit has been performed at each observing spacecraft in the alignment independently. Fit parameters at the inner spacecraft are listed on the left-hand side of Table 4, and parameters at the outer spacecraft on the right-hand side.

The final $\chi^2$ value and the $\delta$ fit uncertainty measure are also given in Table 4 for each fitting. Fits have been judged to be sufficiently accurate when $\chi^2 < 1.5$ and $\delta < 10°$. It can be seen qualitatively that fits satisfying these conditions reproduce observations well. Fits to 13 of the 18 total events satisfied the $\chi^2$ and $\delta$ conditions. For each of the 13 events, the conditions were satisfied by the fits at both observing spacecraft. It was notable that, where a good fit was obtained at one spacecraft, a good fit could also generally be obtained at the other spacecraft. The fits to the Q1 events are shown in Figure 8.

The high field magnitudes of the magnetosheath intervals that have been included (i.e., those in which the flux rope orientations were not obviously perturbed by bow shock crossings) have not biased the estimated flux rope orientations since the orientations were determined with normalized data. Event 16 in Figure 8 shows an example of flux rope rotations remaining intact within a magnetosheath. The determination of $B_0$, in contrast, is sensitive to the field magnitude, and all magnetosheath intervals were removed when determining the $B_0$ values.

## 4.2 Fitting results

We now consider how the fitted parameters varied between the aligned spacecraft. In the following section, parameter subscripts 1 and 2 refer to the inner and outer spacecraft locations, respectively; uncertainty values are given by the standard error of the mean. Figure 9 shows how $\theta_0$, $\varphi_0$, $B_0$ and $p$ varied from one spacecraft to the other for each alignment as a function of heliocentric distance, $r$.

The handedness $H$ was the same at the inner and outer spacecraft for all events. Of the 13 events fitted, 5 were right-handed ($H = +1$) and 8 were left-handed ($H = -1$). The impact parameter $p$ was lower at the outer spacecraft in 9/13 cases (black data points in Figure 9, panel (c)). Most of the $p$ values were low, with 9/13 having $p < 0.25$ at the inner spacecraft and 10/13 with $p < 0.25$ at the outer spacecraft. The requirement of clear flux rope signatures when identifying events for analysis may partly explain the prevalence of low $p$ values: rope signatures tend to be more clear for low-$p$ spacecraft encounters. We also note that fitted axis orientations for high-$p$ encounters tend to have a higher associated error.

The absolute value of $\theta_0$ was lower at the outer spacecraft in 10/13 cases (black data points in Figure 9, panel (a)), which may indicate a tendency for the rope axis to reduce in inclination relative to the solar equatorial plane. The absolute change in inclination, $\Delta\theta_0 = |\theta_{02}| - |\theta_{01}|$, had a small mean value of -8.8 ± 4.3º across all the fitted events. Axis inclinations were generally low, with 6/13 events having an inclination of 20º or less at the inner spacecraft, rising to 9/13 events at the outer spacecraft. The magnitude of $\Delta\theta_0$ showed no significant dependence on the propagation distance, $\Delta R = R_2 - R_1$, or the fractional change in heliocentric distance, $R_2/R_1$.

There was a larger spread in axis longitude angles, $\varphi_0$, at the heliocentric distances of MESSENGER and Venus Express (~ 0.3-0.7 AU) compared to 1 AU. In Figure 9, panel (b), there is some tentative indication that the rope axes tended towards the solar east ($\varphi_0 = 270$º) or solar west ($\varphi_0 = 90$º) directions. The mean absolute change in the axis longitude angle, $|\Delta\varphi_0| = |\varphi_{02} - \varphi_{01}|$, was equal to 42.6 ± 9.4º. We note that the particularly large $\Delta\varphi_0$ value for Event 1 is consistent with the spacecraft having sampled significantly different regions of the flux rope (see Section 2.2.3). As with $\Delta\theta_0$, $|\Delta\varphi_0|$ showed no significant dependence on $\Delta R$ or $R_2/R_1$.

The evolution of $B_0$ with heliocentric distance $r$ [AU] may be described with the power law relation $B_0 = kr^c$, where $k$ [nT AU$^{-c}$] and $c$ are constants. A fit to the 26 ensemble values of $B_0$ listed in Table 4 (with each ICME providing two $B_0$ values) gives $k = 12.5 \pm 3.0$ and $c = -1.76 \pm 0.04$. These parameters were obtained from an unweighted least-squares linear fit to the logarithmic values of $B_0$ and $r$, where error values give the 95% confidence level of the fit. Similar fitting was performed by Leitner et al. (2007), who fitted $B_0$ values obtained from 130 events; for the subset of events observed in the inner heliosphere ($r < 1$ AU), Leitner et al. found fit values of $k = 18.1 \pm 3.8$ and $c = -1.64 \pm 0.40$.

The radial dependence of $B_0$ may also be determined for the ICMEs individually by performing separate power law fits to the two $B_0$ values obtained from each event. These separate fits are shown in Figure 9, panel (d). The mean fit parameters for the 13 fits were $k = 17.3 \pm 12.8$ and $c = -1.34 \pm 0.71$ respectively, where the uncertainty values are the standard deviations. The large standard deviations in $k$ and $c$ reflect the large spread in radial dependencies displayed by the individual events. Farrugia et al. (2005) have previously determined the $B_0$ power-law dependence of individual ICMEs in a similar way, and found a similarly broad spread in the $c$ exponent.

| Event # | Q | s/c 1 | | | | | | | | | s/c 2 | | | | | | | | | $\Delta\theta_0$ [°] | $|\Delta\varphi_0|$ [°] |
|---|---|---|---|---|---|---|---|---|---|---|---|---|---|---|---|---|---|---|---|---|---|
| | | H | $\theta_0$ [°] | $\varphi_0$ [°] | $y_0$ | $B_0$ [nT] | p | D | $\chi^2$ | $\delta$ [°] | H | $\theta_0$ [°] | $\varphi_0$ [°] | $y_0$ | $B_0$ [nT] | p | D | $\chi^2$ | $\delta$ [°] | | |
| 1 | 3 | -1 | 35.5 | 142.7 | -0.16 | 22.1 | 0.12 | 0.77 | 0.11 | 4.1 | -1 | 1.9 | 13.8 | 0.11 | 11.5 | 0.01 | 0.24 | 0.27 | 6.4 | -33.5 | -129.0 |
| 2 | 1 | -1 | -27.3 | 46.2 | 0.13 | 12.4 | 0.07 | 0.79 | 0.39 | 4.6 | -1 | -14.9 | 78.6 | 0.09 | 8.8 | 0.02 | 0.98 | 0.30 | 1.5 | -12.4 | 32.4 |
| 3 | 2 | -1 | -4.7 | 118.3 | 0.07 | 14.1 | 0.01 | 0.88 | 0.43 | 4.7 | -1 | -13.1 | 64.0 | -0.42 | 7.9 | 0.10 | 0.91 | 0.29 | 5.8 | 8.4 | -54.3 |
| 4 | 3 | 1 | -30.6 | 224.6 | 0.74 | 53.7 | 0.46 | 0.89 | 0.10 | 6.4 | 1 | 1.5 | 308.2 | -0.09 | 11.2 | 0.00 | 0.79 | 0.23 | 4.8 | -29.0 | 83.6 |
| 6 | 2 | -1 | 17.9 | 102.3 | -0.13 | 16.2 | 0.04 | 0.98 | 0.29 | 1.0 | -1 | -24.7 | 66.9 | -0.03 | 9.6 | 0.01 | 0.93 | 0.10 | 0.7 | 6.8 | -35.4 |
| 7 | 3 | -1 | -11.0 | 58.7 | 1.20 | 23.4 | 0.26 | 0.89 | 0.07 | 5.8 | -1 | -6.9 | 74.3 | -0.43 | 11.0 | 0.05 | 0.96 | 0.23 | 3.6 | -4.2 | 15.5 |
| 8 | 1 | 1 | -38.8 | 322.8 | 0.28 | 54.6 | 0.22 | 0.80 | 0.07 | 5.5 | 1 | -33.4 | 272.4 | -0.23 | 19.9 | 0.12 | 1.01 | 0.03 | 0.0 | -5.5 | -50.5 |
| 10 | 2 | 1 | 7.0 | 343.5 | 0.11 | 13.0 | 0.04 | 0.31 | 0.18 | 7.6 | 1 | 4.4 | 333.0 | -0.26 | 8.9 | 0.04 | 0.46 | 0.24 | 5.9 | -2.6 | -10.6 |
| 12 | 3 | -1 | 18.1 | 238.6 | 2.33 | 186.1 | 0.80 | 1.46 | 1.20 | 6.6 | -1 | 13.0 | 212.5 | 0.94 | 48.4 | 0.36 | 0.61 | 0.63 | 6.0 | -5.1 | -26.1 |
| 13 | 1 | -1 | -25.3 | 4.7 | -0.54 | 83.7 | 0.52 | 0.51 | 0.06 | 4.3 | -1 | -41.1 | 59.3 | -0.49 | 59.3 | 0.47 | 0.75 | 0.11 | 4.8 | 15.8 | 54.6 |
| 14 | 1 | -1 | 12.7 | 81.4 | 0.51 | 42.6 | 0.11 | 1.00 | 0.22 | 1.8 | -1 | 5.4 | 124.7 | -0.16 | 9.2 | 0.02 | 0.82 | 0.17 | 4.2 | -7.3 | 43.4 |
| 16 | 1 | 1 | 38.8 | 272.4 | 0.01 | 37.3 | 0.01 | 1.00 | 0.10 | 0.0 | 1 | -4.0 | 283.2 | 0.64 | 36.9 | 0.04 | 0.97 | 0.10 | 1.5 | -34.8 | 10.8 |
| 18 | 1 | 1 | 46.8 | 164.6 | -0.07 | 22.4 | 0.06 | 0.75 | 0.08 | 3.9 | 1 | 36.2 | 172.2 | 0.30 | 18.8 | 0.28 | 0.63 | 0.08 | 2.9 | -10.6 | 7.6 |

**Table 4.** Lundquist fit parameters. $H$ is the flux rope handedness ('1' denoting right-handed and '-1' left-handed), $\theta_0$ and $\varphi_0$ are the latitude and longitude angles of the rope axis direction relative to the *x-y* SCEQ plane, respectively, $y_0$ is a fitting parameter related to the distance between the rope axis and spacecraft trajectory in the model *x-y* plane, $B_0$ is the axial field strength, $p$ is the spacecraft impact parameter, $D$ is the ratio of the flux rope diameter to the observed radial width, and $\chi^2$ and $\delta$ are fit quality measures (see text for details). $\Delta\theta_0 = |\theta_{02}| - |\theta_{01}|$ and $|\Delta\varphi_0| = |\varphi_{02} - \varphi_{01}|$, where subscripts '1' and '2' refer to the inner and outer spacecraft values, respectively.

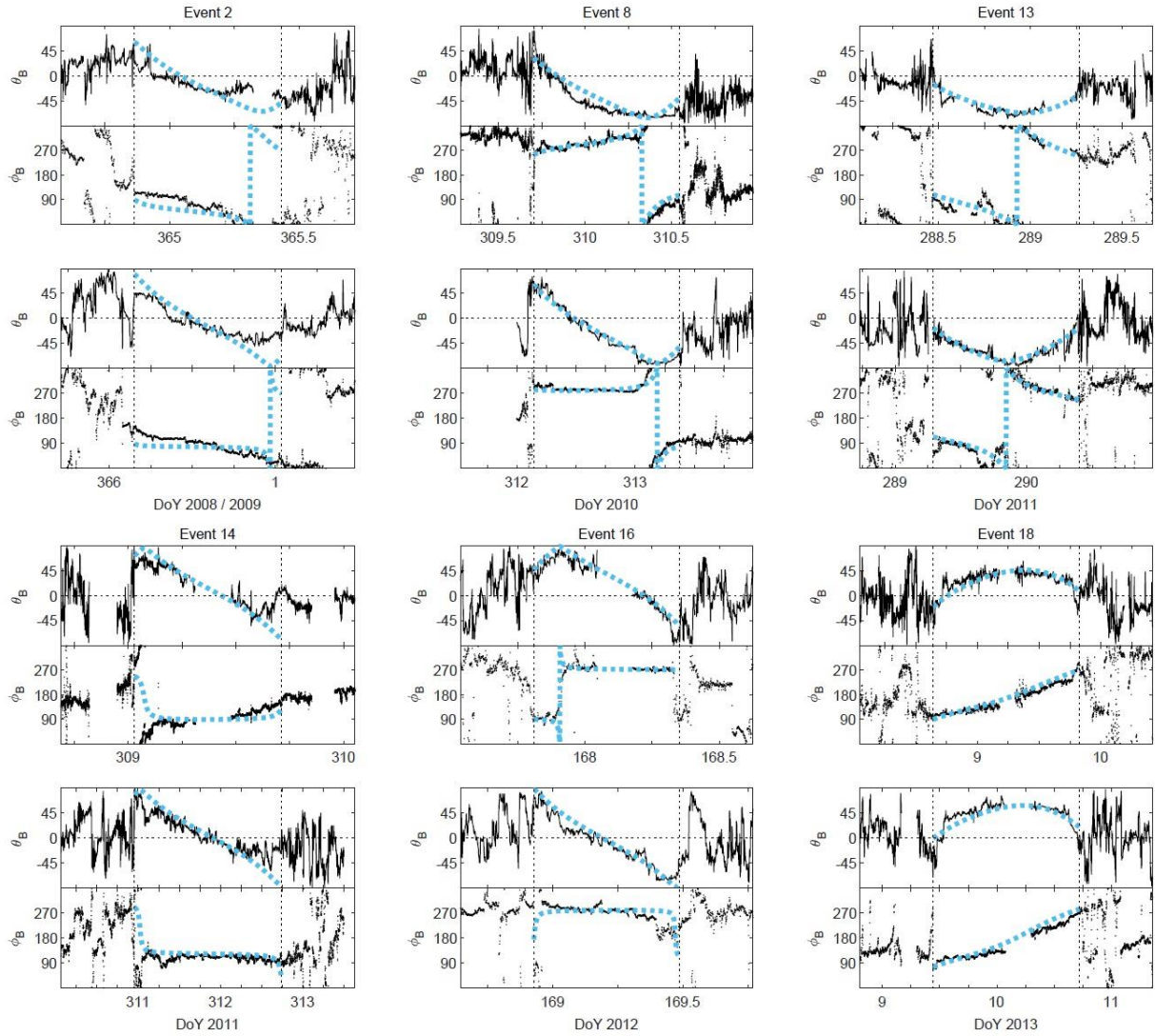

**Figure 8.** Lundquist fits to the Q1 events. The fits are shown by the dashed blue lines overlaying the observed magnetic field direction angles.

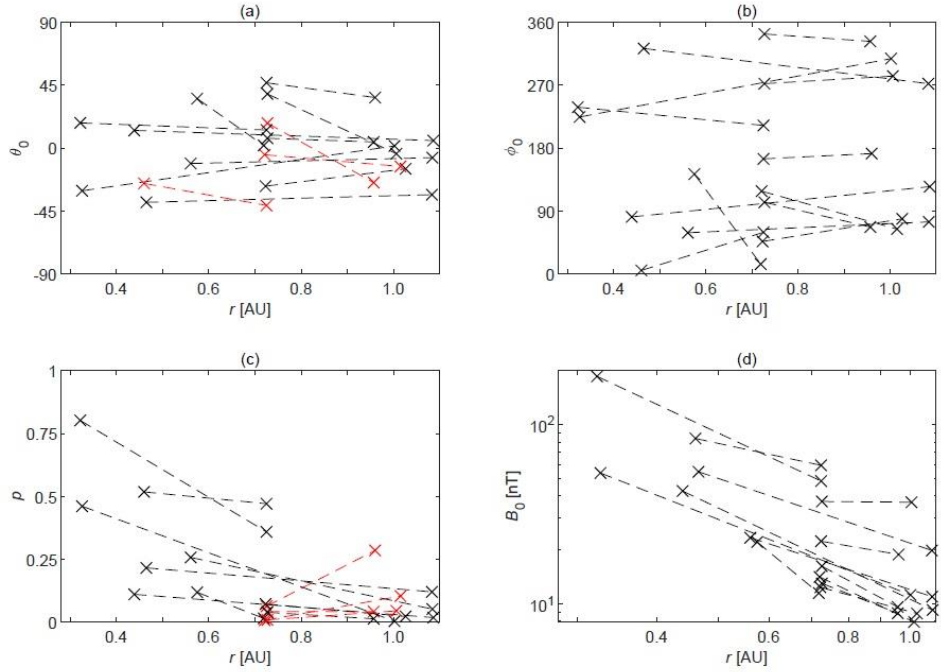

**Figure 9**. Lundquist fit parameters versus heliocentric distance, $r$. The panels show (a) axis direction latitude, (b) axis longitude direction, (c) spacecraft impact parameter, and (d) magnetic field strength at the rope axis. Each rope is fitted independently at two radially separated locations; the dashed lines connect the fit-value pairs for each event. In panel (a), data points for events where the rope axis inclination relative to the solar equatorial plane ($\theta = 0°$) was lower (higher) at the outer spacecraft compared to the inner spacecraft are colored black (red). In panel (c), data points for events where the impact parameter was lower (higher) at the outer spacecraft are colored black (red).

## 5 Discussion

The aim of this work has been to compare the underlying structure of multiple ICME flux ropes observed by pairs of aligned spacecraft at different heliocentric distances in the inner heliosphere. We have defined 'underlying structure' to be the normalized magnetic field time series observed by the spacecraft while inside the flux ropes. Field normalization can remove the effects of certain transient features in the time series (e.g., shock waves) that are not intrinsic to the flux rope field and remove differences in field magnitude between the aligned spacecraft that arise from ICME expansion. A mapping technique has been used to produce figures that overlap the normalized profiles, allowing easy and direct comparison of field features. The flux rope profiles observed at the inner spacecraft are stretched in the time domain to the durations of the ropes at the outer spacecraft.

### 5.1 Rope similarity and orientation

Across the 18 flux ropes analyzed, there were significant similarities in macroscale field structure at the aligned spacecraft. This similarity is evident from qualitative visual inspection of the overlapped data and from measures that quantify differences between the profile magnitudes. We note that the overlap plots reveal similarities in the flux rope field components that are not readily discernible in the unmapped times series (e.g. Event 14) and confirm that the different spacecraft observations were associated with the same ICME.

A higher degree of similarity was generally seen in the Q1 and Q2 mappings, while some of the Q3 mappings showed significant dissimilarities. The greater dissimilarity in the Q3 events suggests that, when ICME flux ropes display more ambiguous signatures or show signs of strong interaction with the solar wind (i.e., Q3 characteristics), then their signatures are less likely to be consistent at a pair of radially aligned spacecraft.

Despite the general similarities, almost all of the events displayed systematic differences in the field direction angle profiles, $\theta_B$ and $\varphi_B$. In the case of the Q1 events, for example, differences ranged from the negligible (Event 13) to the more significant (Event 18). The differences in the $\theta_B$ and $\varphi_B$ profiles are consistent with the flux ropes displaying differing local axis orientations at the aligned spacecraft, as confirmed by the Lundquist fitting analysis.

Figure 10 illustrates how macroscale differences in the profiles can be rotational in nature. In the figure, in which Events 2 and 4 are taken as examples, the rope profile vectors mapped from the inner spacecraft (pale-colored lines) have been rotated such that the rope axis of the mapped profile aligns with the rope axis of the outer spacecraft profile (dark-colored lines). This rotational mapping technique is described in detail by Good et al. (2018). The rotated mappings in Figure 10 show significantly greater similarity than the corresponding mappings in Figures 4 and 6. Comparable increases in profile similarity can be obtained for the other events when aligning the rope axes in this way.

Assuming that the same region of each ICME was sampled by the aligned spacecraft, the Lundquist fitting results presented in Figure 9 suggest a weak tendency for the rope axis to reduce in inclination relative to the solar equatorial plane with propagation distance, and to tend weakly in direction towards the east-west line. There is evidence that ICME flux ropes align with the heliospheric current sheet (e.g., Isavnin et al., 2014, and references therein), which could reduce the rope inclination. The east-west alignment could be due to progressive flattening of the ICME fronts. Statistical distributions of fitted flux rope orientations at 1 AU are consistent with ICMEs having elliptically-shaped fronts (Janvier et al., 2013; Démoulin et al., 45    2016); if an elliptical front flattens normal to the propagation direction over time (i.e., if the ellipse aspect ratio increases with propagation distance), then east-west flux rope axis directions will be observed more frequently at larger heliocentric distances. The prevalence of east-west flux rope axis directions at 1 AU has previously been found in statistical studies of observations (Lepping et al., 2006) and in ICME modeling (e.g., Owens, 2016); east-west alignment is also

50    common in flux ropes observed at sub-1 AU heliocentric distances (Bothmer & Schwenn, 1998; Leitner et al., 2007). The general decrease in impact parameter may also be due to pancaking: as pancaking develops with propagation distance, a wider range of spacecraft trajectories through the ICME may appear like low-impact intersections of a cylindrical rope (Russell & Mulligan, 2002b).

55         Alternatively, the spacecraft could have been sampling different regions of the ICMEs, with the different regions having different local axis orientations. This latter interpretation suggests that ICME properties such as axis orientation can change significantly over the relatively small heliospheric latitudes and longitudes (typically less than 10°) by which the spacecraft were separated. A combination of both effects – temporal evolution and sampling of

60    different parts of a globally curved flux rope  – could also account for the different orientations. Both effects may be required to explain the relatively large mean $|\Delta\varphi_0|$ value of 42.6° obtained in this study. We note that this value is consistent with studies that have found statistical averages of $\varphi_0$ vary significantly with heliocentric distance, particularly in the inner heliosphere (Farrugia et al., 2005; Leitner et al., 2007).

65         Besides highlighting similarities in macroscale flux rope structure, the mappings displayed in Section 3 also reveal some similarities in mesoscale structure, i.e., second-order field features with durations ranging from minutes to hours. An example is highlighted by the gray box overlaying Figure 10; there is a remarkable degree of correlation in this particular field feature given that it was observed ~36 hours apart by widely separated spacecraft. The

70    preservation of these features reflects the stability of the flux rope field in low-$\beta$ plasma. We note, however, that mesoscale similarity is much less prevalent than macroscale similarity in the ICMEs analyzed.
        We have not considered how the field components in a flux rope-centered coordinate system evolve. The axial ($B_A$) and tangential ($B_T$) components may be subject to different

75    physical constraints and evolve with different dependencies on heliocentric distance, as shown by theoretical studies (e.g., Démoulin & Dasso, 2009; Osherovich et al., 1993) and observation (e.g., Russell et al., 2003). The near self-similar evolution of flux rope structure found in the events analyzed in this work could be used to set observational constraints on theoretical models.

## 5.2. Erosion

80         Any effects of magnetic reconnection have not been considered. ICME flux ropes often appear to have been eroded through reconnection by the time they reach 1 AU (Ruffenach et al., 2015). Reconnection with the ambient magnetic field can effectively peel away the outer field lines of the flux rope, reducing the rope's total flux content. Up to two-thirds of the erosive reconnection seen at 1 AU is thought to occur within the orbital distance of Mercury;

85    reconnection rates fall in proportion to the Alfvén speed with increasing heliocentric distance (Lavraud et al., 2014). It is not clear whether a significant amount of erosion would occur at the heliocentric distances over which the ICMEs analyzed in this study propagated, particularly for the events propagating from 0.72 to 1 AU. If it were significant, the rope leading and trailing edges at the inner spacecraft would not map to the rope edges observed at the outer spacecraft, 90     since some of the inner profile (whether at the front or back) would have been eroded away by the time of arrival at the outer spacecraft. It would be worthwhile to investigate whether better overlaps of field features are achieved by reverse-mapping the outer profile edges to internal features of the inner profiles.

95

100

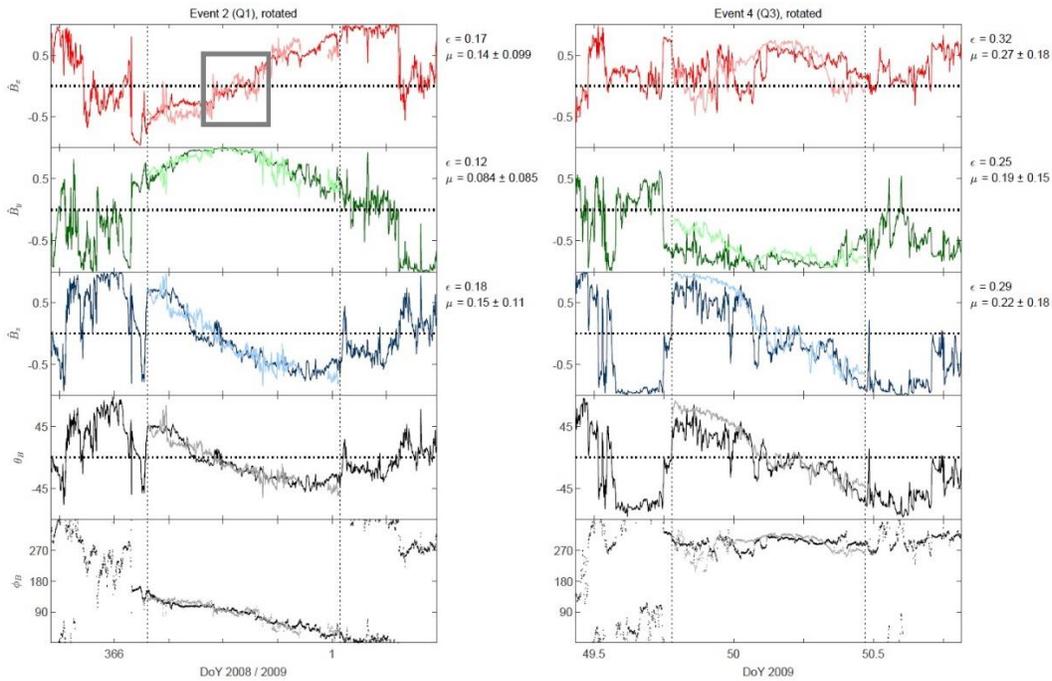

**Figure 10.** Rotated mappings for Events 2 and 4, in which the mapped flux rope profiles have been transformed such that the axis directions of the mapped flux ropes are the same as those of the ropes observed at the outer spacecraft. Gray boxes overlaying the $x$-component profile for Event 2 and the $y$-component profile for Event 16 highlight examples of similar mesoscale features. (Same format as Figure 4.)

110

115

## 6 Summary and Conclusion

120  We have analyzed 18 ICME flux ropes observed by pairs of radially aligned spacecraft in the inner heliosphere. A magnetic field overlapping technique has been used to show the general underlying similarity of the flux rope profiles at the aligned spacecraft. This similarity was seen across a range of spacecraft separation distances. As well as revealing similarities in macroscale structure, the overlapping has also shown how mesoscale field features were sometimes observed

125  at both spacecraft.

A preliminary analysis of the ICMEs' kinematic behavior has been performed. Mean transit speeds of the flux rope leading edge and center of mass were generally similar to the speeds observed at 1 AU, while trailing edge speeds tended to increase. Increasing trailing edge speeds were correlated with relatively fast solar wind to the rear of the ICMEs. The ICME

130  expansion speeds tended to decrease or remain constant; reductions in the expansion speed were generally caused by increases in the trailing edge speed.

The global structure and orientation of the flux ropes has been characterized with Lundquist fitting. Where there were macroscale differences in the profiles – in most cases minor, in some cases more significant – these differences may be ascribed to differences in the

135  flux rope axis orientation. Orientation changes may haven be due to the flux ropes aligning with the heliospheric current sheet and flattening of the ICME front with distance from Sun. Alternatively, the differences may have been due to the spacecraft sampling different regions of the flux rope, with the different regions having different axis orientations. Since the spacecraft angular separations were small on heliospheric scales, the latter interpretation suggests that

140  global properties of ICMEs such as axis orientation can vary significantly across small angular distances.

The 18 events analyzed in this study represent the clearest examples identified by Good and Forsyth (2016) where radially aligned spacecraft both observed flux rope signatures. In their analysis of more than 100 ICMEs observed by MESSENGER and Venus Express, Good and

145  Forsyth found that when a spacecraft observed flux rope signatures, a second spacecraft at a greater heliocentric distance and separated by less than 15° in heliographic longitude subsequently observed flux rope signatures in 82% of cases; the present study has demonstrated that the flux rope signatures are likely to be similar at the two spacecraft. This finding supports the case for an upstream space weather monitor sitting on or near the Sun-Earth line: the first-

150  order flux rope structure observed at such a monitor (and the normalized $B_z$ component of that structure) is likely to be the same as that arriving subsequently at the Earth, even in cases where the radial separation distance between the monitor and Earth is large (e.g., Event 8). With an estimation of how the field magnitude profile also evolves (as recently considered by Janvier et al., 2019), simple and accurate $B_z$ forecasting with a near-Sun upstream monitor may be

155  possible. However, differences due, for example, to a change in rope orientation may be difficult to predict without global modeling. Furthermore, the 1 AU signatures of a significant minority of ICMEs – e.g., the 18% of events identified by Good and Forsyth that did not display flux rope signatures at a second, aligned spacecraft, and the complex ICME reported by Winslow et al. (2016) – would not be easily predicted solely with observations from an upstream monitor.

160  **Acknowledgements**

We wish to thank the two anonymous reviewers for their thoughtful consideration of the manuscript. Data used in this work was obtained from the ICMECAT catalog, a product of the HELCATS project; the archived catalog data may be found at https://doi.org/10.6084/m9.figshare.4588315.v1. Data included in Table 1 may be found in an
165   easily downloadable form at https://doi.org/10.6084/m9.figshare.8143568.v1. S.G. and E.K. are
supported by ERC Consolidator grant ERC-COG 724391 (SolMAG), and by Academy of
Finland grants 310445 (SMASH) and 312390 (FORESAIL). This work has also been supported
by the European Union Seventh Framework Programme under grant agreement No. 606692
(HELCATS). C.M. thanks the Austrian Science Fund (FWF): [P31521-N27]. We also wish to
170   thank the MESSENGER, Venus Express, STEREO and Wind instrument teams.